# L-Band Milliwatt Room-Temperature Solid-State Maser

Sophia Long[1], Max Attwood[2], Justin Chan[2], Priyanka Choubey[1], Hamdi Torun[1], Juna Sathian[1]

*[1]School of Engineering, Physics and Mathematics, Faculty of Science and Environment, Northumbria University, NE1 8ST, United Kingdom.*

*[2]Department of Materials, Imperial College London, Exhibition Road, London, SW7 2AZ, United Kingdom.*

## Abstract

Molecular room-temperature masers have emerged as promising sources of coherent microwaves, but systematic comparisons of available organic gain media under uniform conditions have remained elusive. This paper presents a comprehensive characterisation of two established organic gain media, pentacene doped para-terphenyl (Pc:PTP, 1.45 GHz) and 6,13 diazapentacene doped para-terphenyl (DAP:PTP, 1.478 GHz), studied at four concentrations, three previously reported (0.1% and 0.01% Pc:PTP; 0.01% DAP:PTP) and one newly introduced in this work (0.05% DAP:PTP). This work addresses a previously under-explored aspect by evaluating reported L-band masing samples under identical conditions, providing insights into the optimal concentrations and gain media for specific applications. An optimised system produced a room-temperature milliwatt maser with a peak power of 2.34 mW (+3.69 dBm), marking a milestone in achieving whole-milliwatt output and demonstrating potential for high-power applications. Spectral coherence times of 465 ns and coherence lengths reaching 150 m were achieved with this setup. Coupling to the high-Q $TE_{01\delta}$ cavity mode enables collective spin-photon interactions, where time-domain Rabi oscillations reflect coherent Dicke state dynamics of the spin ensemble. In the frequency domain, FFT analysis reveals normal-mode splitting of 1.37 MHz for Pc:PTP, and 2.14 MHz for DAP:PTP, confirming the system operates in the strong-coupling regime. Strong coupling is corroborated by cavity-QED analysis, which revealed cooperativities of $C^* \in [304, 803]$ for Pc:PTP and $C^* \in [405, 1071]$ for DAP:PTP, among the highest reported for organic maser systems. The feasibility of applying these coherent masers to advanced technologies such as radar, secure communications, and qubit interfaces is assessed quantitatively by metrics of signal-to-noise ratio (SNR), spectral coherence distance (SCD) and throughput.

## Introduction

Coherence is a fundamental property of electromagnetic waves and a key feature in radio frequency (RF) technologies, where phase relationships are maintained over extended distances and timescales to enable precise signal transmission and interference control. Understanding coherence in RF waves involves analysing their temporal and spatial phase consistency, directly affecting phase stability, synchronisation, and overall system performance [1]. In this study, coherence is examined in both the time and frequency domains, focusing on temporal coherence of spin-photon interactions and frequency-domain normal-mode splitting as key performance indicators of maser operation and quantum electrodynamics. The ability to maintain coherence underpins a wide range of applications, from timekeeping frequency standards and wireless communication to radar and quantum sensing, where preserving phase stability is essential for achieving high precision and optimal performance [2], [3]. In particular, the L-band (1-2 GHz) is central in modern RF technology, serving as a foundation for satellite communication, navigation (e.g. GPS/GNSS) [4] [5], and maritime systems [6]. Its combination of long-range propagation, resilience to atmospheric attenuation (clouds, rain, foliage) [7], and compatibility with low-power terminals [8] [9] makes it highly valued for reliable links in challenging environments.

The maser, an early precursor to the laser, is a device that utilises stimulated emission to amplify RF signals in the GHz range with minimal added noise. Since the invention of the maser in the 1950s, maser technology has played a fundamental role in precision science. Early milestones included the development of the ruby maser, which served as a low-noise amplifier for microwave signals. However, it suffered from the need for cryogenic cooling ($\lesssim$4.2 K [10]) and vacuum conditions, limiting its practicality, reproducibility, and scalable output power. As a result, maser amplifiers were eventually superseded in many applications by solid-state technologies such as Monolithic Microwave Integrated Circuit (MMICs) despite their generally higher noise levels [11].

There has been a resurgence in maser technology since the first demonstration of a room-temperature maser [12]. With 200 W of optical pump power, an organic mixed molecular crystal of p-terphenyl doped with pentacene (Pc:PTP, 0.01% molar concentration) produced a 0.1 $mW$ (-10 dBm) burst of microwaves, $10^8$ times greater than the hydrogen maser. Subsequent efforts [13] focused on extending the maser lifetime duration ($\tau$) via an invasive Ce:YAG luminescent concentrator pump design to generate a quasi-continuous 4 ms maser output (Pc:PTP, 0.1% molar concentration) with a peak power of 3.16 $\mu W$ (-25 dBm) [14]. To date, the highest output power for a maser pulse is 0.33 mW ($-5$ dBm) using a diode-pumped pulsed 532 nm Nd:YAG laser at 32 mJ ,with a 0.1% Pc:PTP gain medium [15].

Room-temperature masers face persistent challenges that restrict their wider adoption. Chief among these is the very high optical pump thresholds needed to achieve population inversion, and the difficulty of sustaining sufficiently high gain to produce practical output powers. High gain is not just a performance metric; it is essential for delivering the low-noise operation that makes masers attractive in real-world applications. Without materials capable of combining low thresholds, long coherence times, and efficient spin dynamics, masers remain underpowered despite their exceptional low-noise potential.

Building on this context, this paper presents a comprehensive characterisation of four known pentacene-doped gain media, specifically Pc:PTP (1.45 GHz) and DAP:PTP (1.478 GHz), at varying concentrations (0.1%, 0.01% and 0.05%, 0.01% respectively) under uniform masing conditions.

Among these, the 0.1% Pc:PTP concentration is the most commonly used L-band gain media [16], [17], [18], serving as a reference for performance comparisons. In contrast, 0.01% Pc:PTP has been used fewer times [14], [12], [19], 0.01% DAP:PTP has been reported just once for masing [20], and 0.05% DAP:PTP has not been mased with prior to this study. Here, we demonstrate that low-concentration DAP:PTP exhibits superior maser performance compared with conventional Pc:PTP, linked to a combination of factors, including enhanced optical penetration depth at lower dopant loadings and reduced thermal load during pumping for DAP samples.

## Results

### 3.1 Optical Characterisation

Ultraviolet-Visible (UV-Vis) absorption spectroscopy was employed to gain insights into the optical properties and electronic transitions of the synthesised Pc:PTP and DAP:PTP samples. The recorded spectra (Fig. 1a) span 400 to 700 nm, with a vertical marker at 532 nm, indicating the laser pump wavelength. Notably, the Franck–Condon vibronic bands are observed to have a 30 nm redshift in DAP:PTP compared to Pc:PTP. The redshifted vibronic absorption band of DAP:PTP aligns the 532 nm pump with a strong absorption vibronic transition, whereas in Pc:PTP, the same wavelength falls on the shoulder of the absorption wing, reducing absorption efficiency.

Recent pulsed ODMR measurements indicate that the triplet sublevel populations in DAP:PTP ($P_x$: $P_y$: $P_z$: $\simeq$ 0.74 : 0.14 : 0.11) [21], are similar to those in Pc:PTP (0.76 : 0.16 : 0.08) [22]. While the absorption profiles of Pc:PTP and DAP:PTP differ slightly, both exhibit vibronic features within the 520–540 nm range, near the 532 nm excitation wavelength.

At this wavelength, absorption strength also sets the optical penetration depth. Takeda *et al.* [23] report that 0.1 mol% dopant concentrations restrict penetration to ~0.7–1.0 mm, whereas 0.01 mol% samples allow >2 mm. Mapping these values to a 3 mm pump path (the diameter of the gain media due to transverse pumping) indicates that high-concentration Pc:PTP absorbs $\geq$ 95% of incident light within the first millimetre, while low-concentration DAP:PTP absorbs $\leq$ 78%, transmitting $\geq$ 22% deeper into the crystal. This difference suggests that DAP:PTP at low concentration benefits from more uniform excitation through the bulk, a factor relevant to the maser performance discussed later.

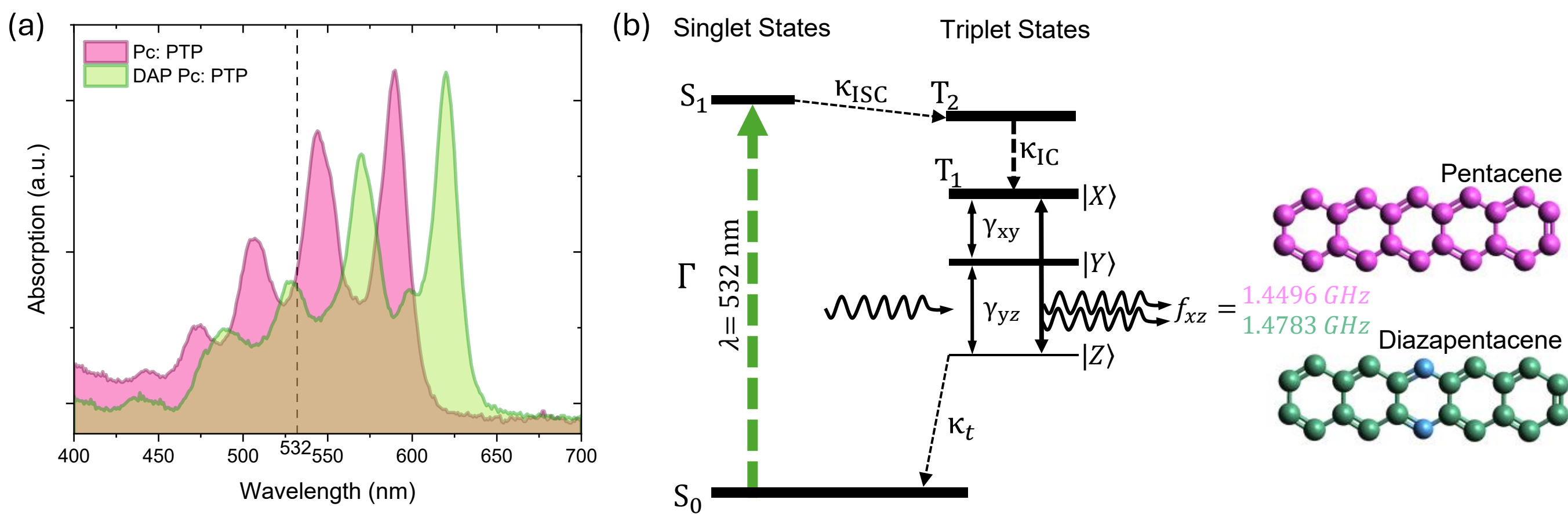

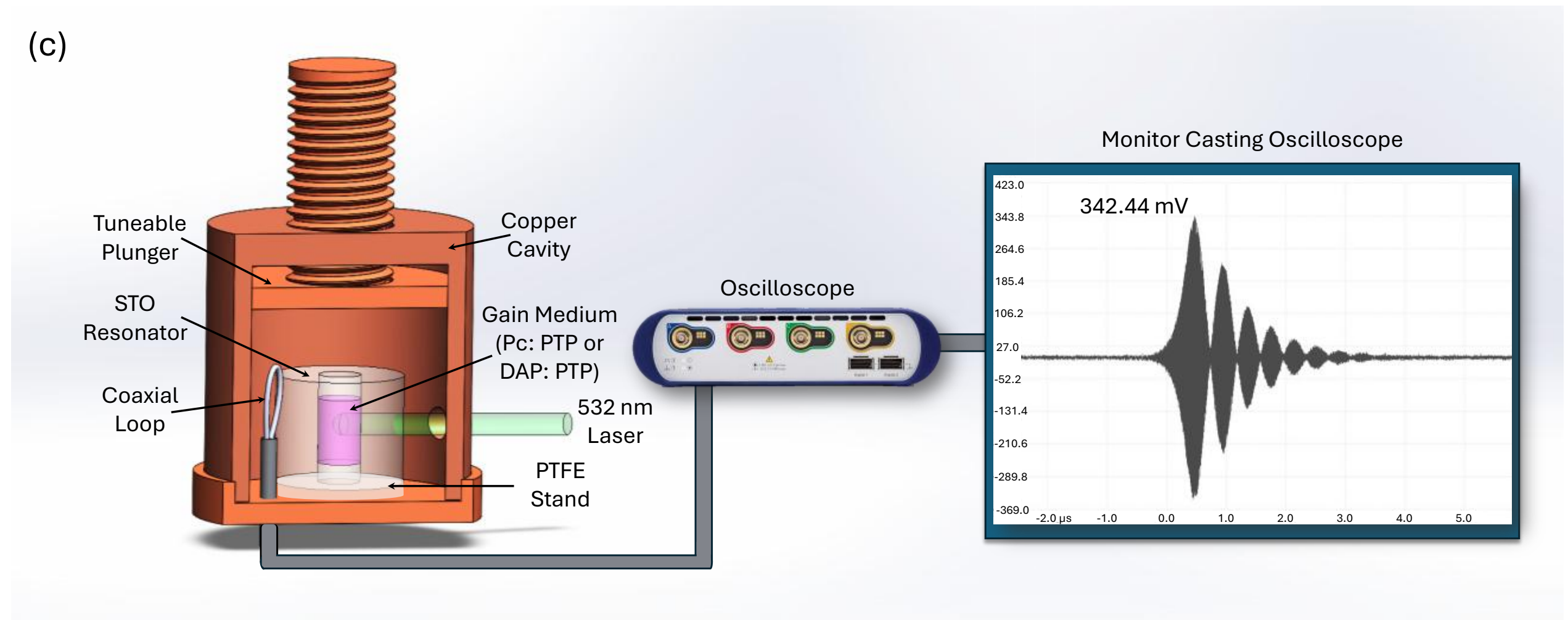

Figure 1: Characterisation and experimental representation of Pc:PTP and DAP:PTP Masers. (a) UV/vis absorption spectroscopy of Pc:PTP and DAP:PTP gain media showing absorption features, the dashed vertical line indicates the 532 nm pump laser wavelength; (b) Jablonski diagram of the state cycle responsible for generating spin polarised triplet-state sublevels in Pc:PTP and DAP:PTP samples under zero-applied magnetic field; (c) Illustration rendering of the maser experiment setup, illustrating the generation and detection of the maser signal. The right-hand panel shows a representative maser pulse with a maximum output power 342 mV observed for the 0.01% DAP:PTP gain medium, consistent with the results discussed in Figure 2a.

The resulting microwave maser signal was extracted via the coaxial coupling loop, which was interfaced with the $TE_{01\delta}$ mode field distribution. This signal was directed to a high-bandwidth oscilloscope (Picoscope 6428E, Pico Technology, Cambridgeshire, UK) for time-domain monitoring.

3.2 Maser Output Mapping

With the optical response established and the 532 nm pump justified, we now quantify how output scales with pump energy at resonant frequency of the maser system, $f_0$. Understanding the behaviour of masing gain crystals under varying optical excitation energies provides valuable insights into the dynamics of room-temperature quantum coherence characteristics. We evaluated the maser output as a function of optical pump energy and tuning frequency. The laser pulse energy was systematically varied from the minimum threshold required for maser action ($P_{min}$= $M_{threshold}$) to the maximum applied energy ($P_{max}$) and was the only variable.

Figure 2 presents a series of graphs mapping the maser output across a laser input energy range from 2.5 mJ to 32 mJ.

Maser thresholds are observed across all samples, with 0.01% Pc:PTP achieving the lowest value (2.5 mJ), whereas DAP:PTP required 5.3 mJ of optical energy under 532 nm pumping. Second, maximum maser output powers are higher at the low-concentration devises, with peak outputs of of 2.34 mW (0.01% DAP:PTP) and 1.79 mW (Pc:PTP).

Finally, the output scaling behaviour showed clear concentration dependence: the 0.01% DAP:PTP sample maintained a closely linear response across the full pump range, whereas the 0.01% Pc:PTP sample exhibited partial saturation at higher pump energies, a two-step regime, and the 0.1% Pc:PTP, 0.05% masers saturated fully.

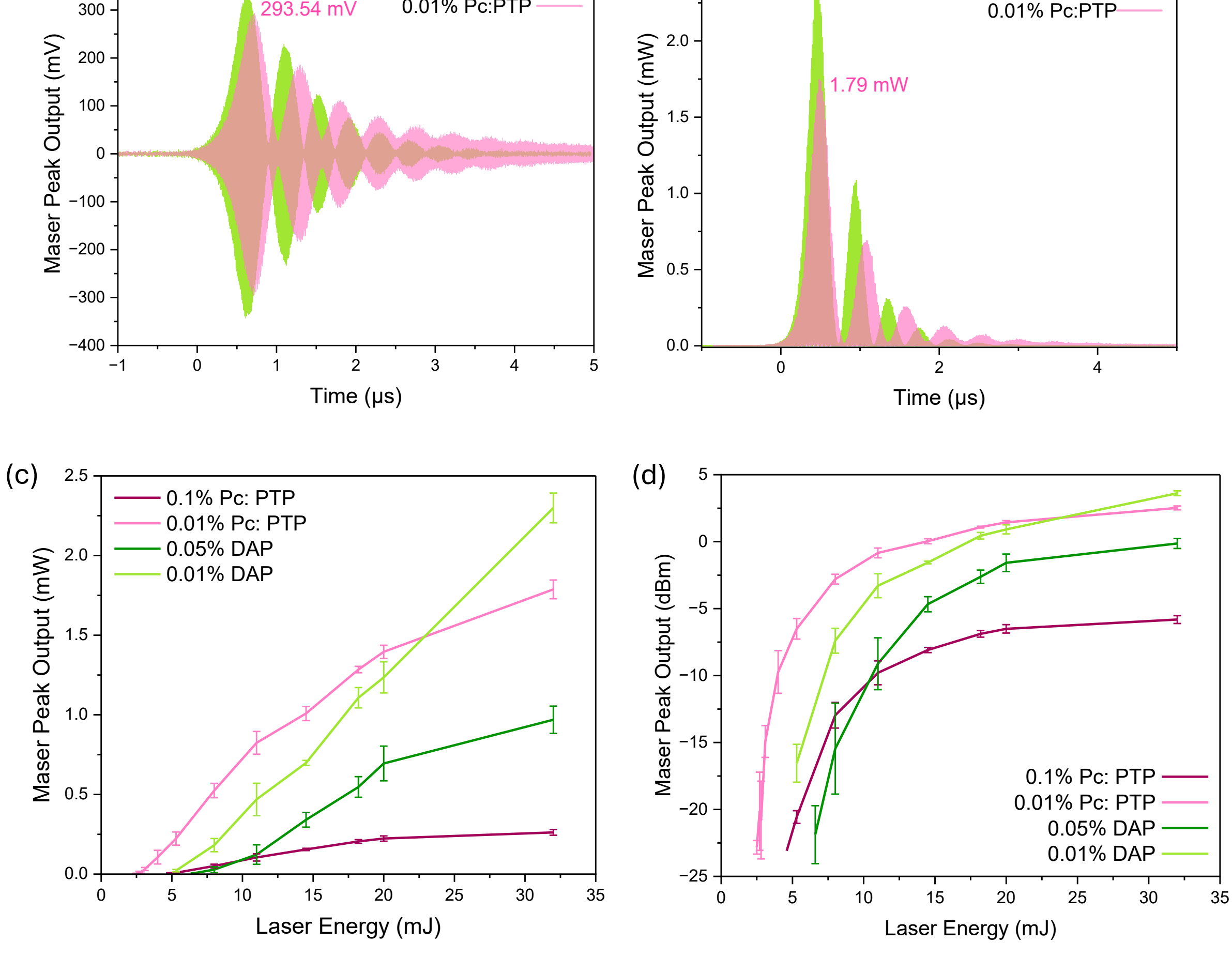

Figure 2: Maser output under varying laser output energy (2.5 mJ to 32 mJ), with the cavity tuned to each sample's respective $f_0$. The variation of maser peak power is calculated using 5-10 datasets. (a) Example time-domain maser signal from the 0.01% DAP:PTP sample demonstrating damped Rabi oscillations with a peak output of 342.44 mV (b) Maser power (mW) vs time ($\mu s$) for the 0.01% DAP:PTP sample at 32 mJ. (c) Maser output as a function of laser beams peak power (mJ), which shows measurement uncertainties for the gain media in the maser output mapping experiment. (d) Maser output in dBm.

The higher maser gain observed in the lower-dopant samples (0.01%) may initially appear counterintuitive, since cooperativity in QED is expected to scale with spin concentration ($C \propto N$). Reducing the dopant concentration from 0.1% to 0.01% decreases the effective spin filling factor by an order of magnitude, limiting the fraction of cavity mode volume occupied by active spins. Despite this, several interrelated physical mechanisms outweigh the reduction in active spins. At higher dopant densities, increased optical absorption reduces the pump light's penetration depth, confining the excitation to the surface and reducing population inversion deeper into the crystal. As reported by Takeda *et al.* [23], ~0.1 mol% dopant limits the excitation depth to ~0.7–1.0 mm, compared with 0.01%, where the reduced absorption allows significantly deeper penetration (>2 mm), enabling more uniform excitation throughout the crystal volume.

Moreover, the increased probability of reabsorption, excited-state annihilation, and localised heating, which in turn degrade the effective cavity quality factor (Q) and induce frequency drift via thermal shifts in spin transition energies. This thermal instability was also highlighted in the review by Breeze *et al.* [24], noting the challenges of balancing pump power, and dopant concentration for continuous maser operation.

Conversely, lower concentrations result in improved pump homogeneity, reduced thermal load, and potentially enhanced spin coherence ($T_2$), which can improve maser performance even when spin number ($N$) is lower. These competing effects suggest that while cooperativity $C \propto N$, optimal maser performance may not correspond to maximum dopant density, especially at room-temperature, optically pumped systems.

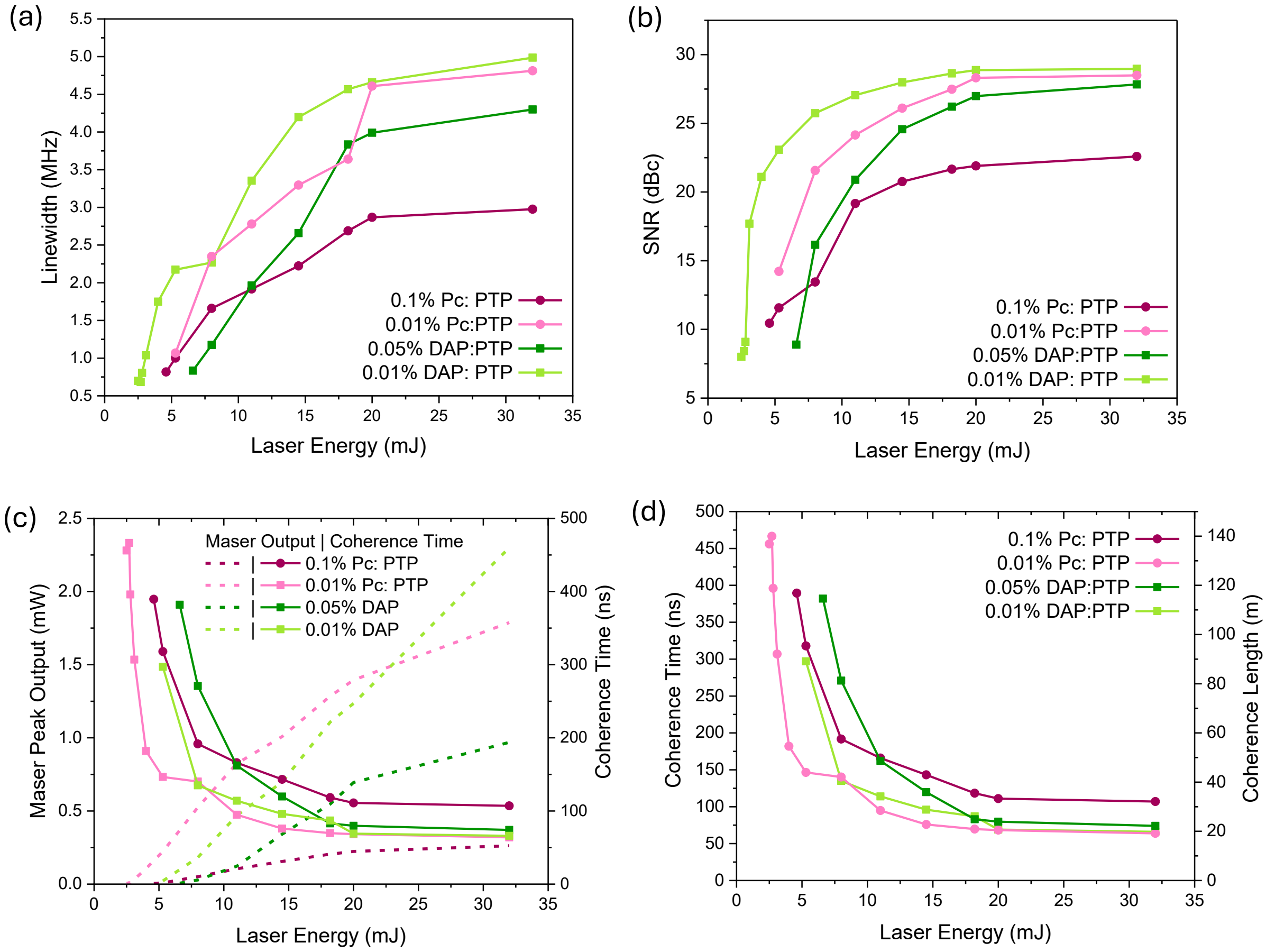

Figure 3: (a) Linewidth as a function of laser energy for all gain media. (b) Signal-to-noise ratio (SNR) vs laser input power, 1 Hz optical pump. (c) Maser output and spectral coherence time vs. laser energy for all samples. (d) Spectral coherence time and coherence length as a function of laser pulse energy.

The maser linewidth shown in Figure 3a was obtained by measuring the full width at half maximum (FWHM) of the maser carrier peak in the frequency domain at each laser energy. The spectrum was acquired using the oscilloscopes internal FFT function, resolving the dominant masing frequency. Linewidths increased with pump energy across all samples, with high-concentration media reaching saturation at lower values than the low-concentration samples.

Figure 3b presents each sample's signal-to-noise ratio (SNR), obtained directly from the PicoScope FFT output and expressed in dBc relative to the carrier. Here, the maser carrier serves as the reference against which broadband noise is measured (see methods). Three samples, 0.01% DAP:PTP, 0.01% Pc:PTP, and 0.05% DAP:PTP, exhibited values nearing 30 dBc. For context, earlier work [18] reported a SNR of 133 (~20 dBc) from a 0.1% Pc:PTP maser. These measurements establish that reducing the dopant concentration yields markedly higher SNR performance than the 0.1% Pc:PTP benchmark.

Figure 3c plots maser peak output against spectral coherence time, showing the expected inverse scaling with increasing pump energy. The 0.01% Pc:PTP sample achieved the longest measured coherence time of 465 ns, while higher-concentration samples showed lessened values.

Figure 3d presents the corresponding spectral coherence lengths, which follow the same trend, extending to ~140 m in the 0.01% Pc:PTP case.

### 3.3 Maser Cavity Frequency Sweep

Having mapped output-pump behaviour at $f_0$, we next swept the cavity to examine frequency dependence and linewidth across the tuning range. To understand how the masing crystals perform across the frequency spectrum, the copper cavity height was adjusted to approximately match the known $f_0$ for each sample using the plunger. Optical excitation was initiated by a 532 nm laser (1 Hz, 10 mJ per pulse, 5 ns pulse duration). The cavity height was the only changing parameter in this experiment, with all other conditions remaining constant.

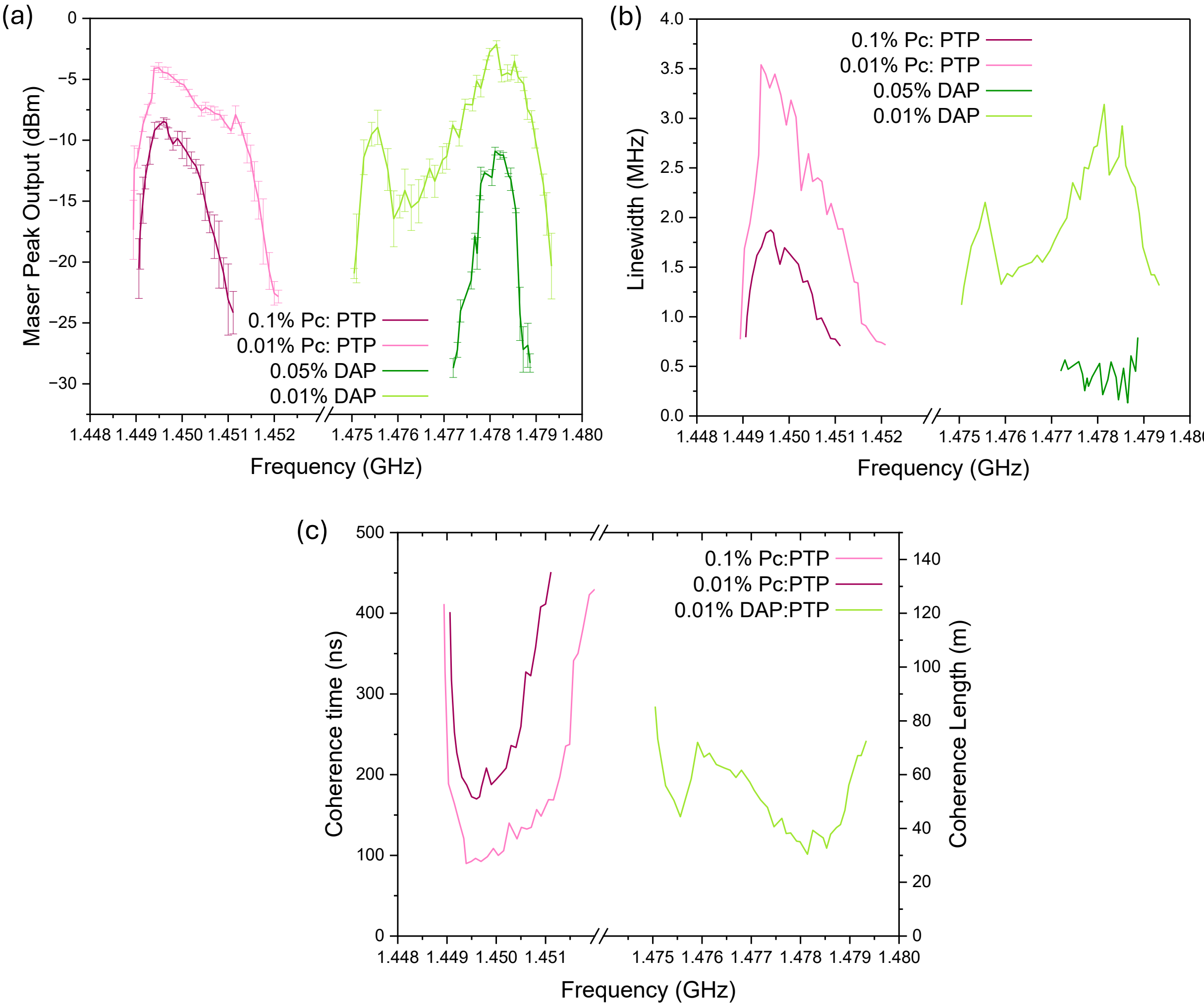

Figure 4: Frequency sweeps derived from the 1 Hz, 10 mJ, optical pump experiment. Each data point represents an average of 5 to 10 microwave signals. Error bars are shown where visible; for (b) and (c), uncertainties were typically <5% and omitted for clarity. (a) Maser peak output frequency sweep presenting the frequency-dependent maser output (dBm) for four gain

mediums (0.1%, 0.01% Pc:PTP, 0.05%, 0.01% DAP:PTP) with error bars across the L-band frequency range (1.448–1.480 GHz). b) Corresponding average linewidth analysis for each gain medium from 3c) Maser peak output frequency sweep presenting the frequency-dependent maser output (mV) for four gain mediums (0.1% Pc:PTP, 0.01% Pc:PTP, 0.05% DAP:PTP, 0.01% DAP:PTP)

The maser frequency-dependent peak output exhibit line shapes consistent with the zero-field ODMR transitions of the respective gain media [25]. The observed frequency sweeps (Fig. 4a) reflect the intrinsic emission characteristics of the $X \leftrightarrow Z$ triplet transitions in both Pc:PTP and DAP:PTP [21]. The pre-$f_0$ shoulder in the DAP:PTP maser linewidth wad attributable to hyperfine coupling to the two $N^{14}$ spins (I = 1), while the asymmetric features in Pc:PTP arise from second-order hyperfine interactions between the triplet spins and pentacene protons.

Across the tuning range, higher dopant concentrations produced lower peak maser output and reduced tuning span (Fig. 4). This trend is also reflected in the maser spectral linewidth. The 0.05% DAP:PTP sample was noticeably less stable, displaying monotonic linewidth changes with pump energy and irregular coherence times, making results less repeatable than the other media, detailed in the Supplementary Information (Fig. S2).

Linewidth and coherence time follow the expected inverse relationship (Fig. 4d). Narrower linewidths correspond to a longer coherence time, defining a microwave field with stable, predictable phase. This phase stability sets the spectral coherence length, with the 0.01% Pc:PTP sample reaching $\sim 465$ nm ($\sim 150$ m). Beyond this length $L_c$, the signal amplitude persists but its phase becomes uncertain, limiting its utility in phase-sensitive applications. Thus, $\tau_c$ and $L_c$ serve as benchmarks for maser performance under varying gain and power.

At the same time, higher microwave output power, while beneficial for signal strength, broaden the linewidth due to gain saturation and increased phase noise. Consequently, coherence times decrease at higher pump energies, underscoring the trade-off between maximising output and preserving phase stability. This balance is consequential to optimising masers for coherence-sensitive fields.

### 3.4 Maser and cQED Experiments

Using the tuning curves and linewidth behaviour, we analyse emission spectra for normal-mode splitting and derive bounds on coupling and cooperativity. To assess whether Pc:PTP and DAP:PTP operate within the strong coupling regime required for cavity quantum electrodynamics (cQED) necessary for quantum interfaces, we evaluate the system cooperativity, $C^*$, defined as $C^* = 4g_e^2/\kappa_c\kappa_s$. Strong coupling is achieved when the ensemble spin–photon interaction exceeds both the cavity decay rate $\kappa_c$ and the spin decoherence rate $\kappa_s$, such that $C^* > 1$.

In this study, $g_e$ requires detailed cavity-QED simulations or fit models [20] [26]. We adopt values for $g_e$ and $\kappa_s$ directly from ref [20] which previously used 0.01% DAP:PTP. For spin–photon coupling strength, we use, $g_{e,sim} = 2\pi \times 2.3\, MHz$. The spin decoherence rate $\kappa_s$, is determined by the inhomogeneous dephasing time of $T_2^*$. Following prior work, we take $\kappa_s = 2/T_2^*$. Examples of reported values of $T_2^*$ span from 1.1 μs [20] for DAP:PTP to 2.9 μs for Pc:PTP samples [26]. Assuming that these values are consistent between like-for-like concentration samples of each spin system, we have estimated $C^*$ across this range. The cavity decay rate, $\kappa_c$, calculated as $\kappa_c = \omega_c \,/\, Q$, with $\omega_c = 2\pi \times f_c$, where $f_c$ is the cavity resonant frequency and Q is the loaded quality factor of the resonator. For Pc:PTP and DAP:PTP, we find $\kappa_c = 2\pi \times 0.24\, MHz$ and $\kappa_c = 2\pi \times 0.18\, MHz$, respectively. This yields a cooperativity of $C^* \in [304 - 803]$ for Pc:PTP and $C^* \in [405 - 1071]$ for DAP:PTP. For literature comparison, in the 1.45-1.47 GHz band, and $\kappa_s$ (1.1 μs − 2.9 μs) reported room-temperature values, Breeze *et al*. reported $C^* {\sim} 190 - 250$ (Pc:PTP), while Ng *et al*. report $C^* = 182$ (DAP:PTP). This translates to Pc:PTP cooperativities $1.6 - 4.2 \times$ (or $1.22 - 3.21 \times$) higher than those reported by Breeze *et al.*, and DAP:PTP values are $2.2 - 5.9 \times$ higher than those reported by Ng *et al.*

As shown in Figure 5, the Pc:PTP sample exhibits a normal-mode splitting of 1.37 MHz, while the DAP:PTP sample yields a larger splitting of 2.14 MHz at 32 mJ of optical pumping. Using the relation $g = \Delta f / 2$ these values correspond to coupling strengths of 0.69 MHz and 1.07 MHz respectively, confirming stronger collective coupling in DAP:PTP.

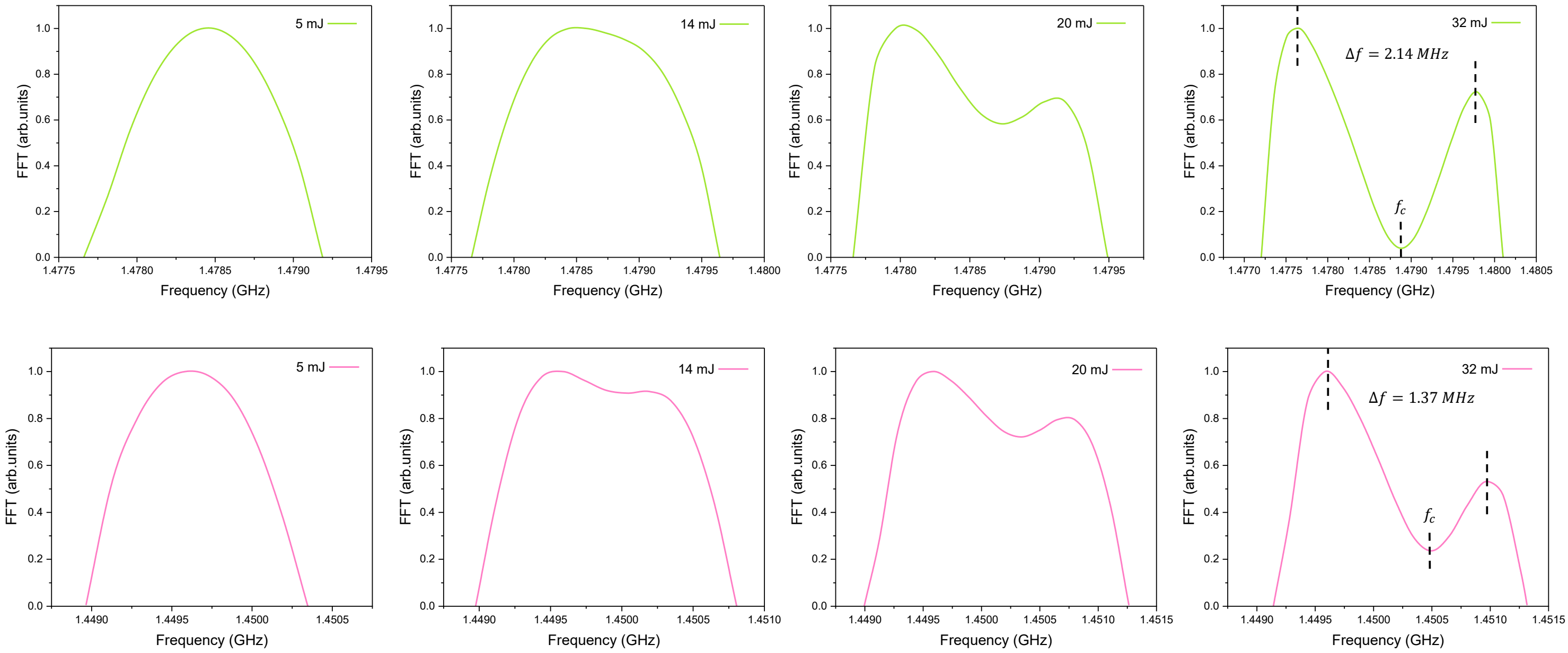

Figure 5: Fourier analysis of the maser output for 0.01% Pc:PTP and 0.01% DAP:PTP at increasing pump energies of 5 mJ, 14 mJ, 20 mJ, and 32 mJ. At low pump energies (5 mJ, 14 mJ), the spectra show single-peaked responses, while at higher energies (20 mJ, 32 mJ) normal-mode splitting emerges, with measured frequency separations of $\Delta f = 2.14$ MHz for DAP:PTP and $\Delta f = 1.37$ MHz for Pc:PTP. We also extracted Rabi frequencies from the time-domain maser transients by measuring envelope peak-to-peak periods ($f_R = 1/T_{peak}$) (see table 1). Pc:PTP yielded $f_R \approx 2.0\, MHz$ compared to $\Delta f = 1.37$ MHz from FFT, and DAP:PTP yielded $f_R \approx 2.5\, MHz$ compared to $\Delta f = 2.14\, MHz$. These values are consistent within experimental resolution, with small offsets attributable to sampling rate and damping.

To further analyse the coupling dynamics in these systems, we consider the normal-mode splitting behaviour that arises under the Tavis-Cummings framework. This model describes the interaction between a single cavity mode and an ensemble of $N$ two-level spins and predicts that the spin-photon coupling strength scales as $g_e\sqrt{N}$. The resulting coherent hybridisation of the spin ensemble with the cavity field leads to energy-level splitting observable in the masing frequency spectrum, referred to as normal-mode splitting. In the masing regime, where stimulated emission dominates, the photon population within the cavity grows substantially and can approach $n_{cavity} \gg 1$. Under such conditions, the electromagnetic field begins to behave classically, and the system departs from the idealised vacuum Rabi regime. This saturation effect is reflected in the narrowing of frequency-domain splitting compared to the ideal Rabi frequency $\Omega_{rabi}$, and in some cases the splitting becomes proportional to $\sqrt{n_{cavity}}$ rather than $g_e$.

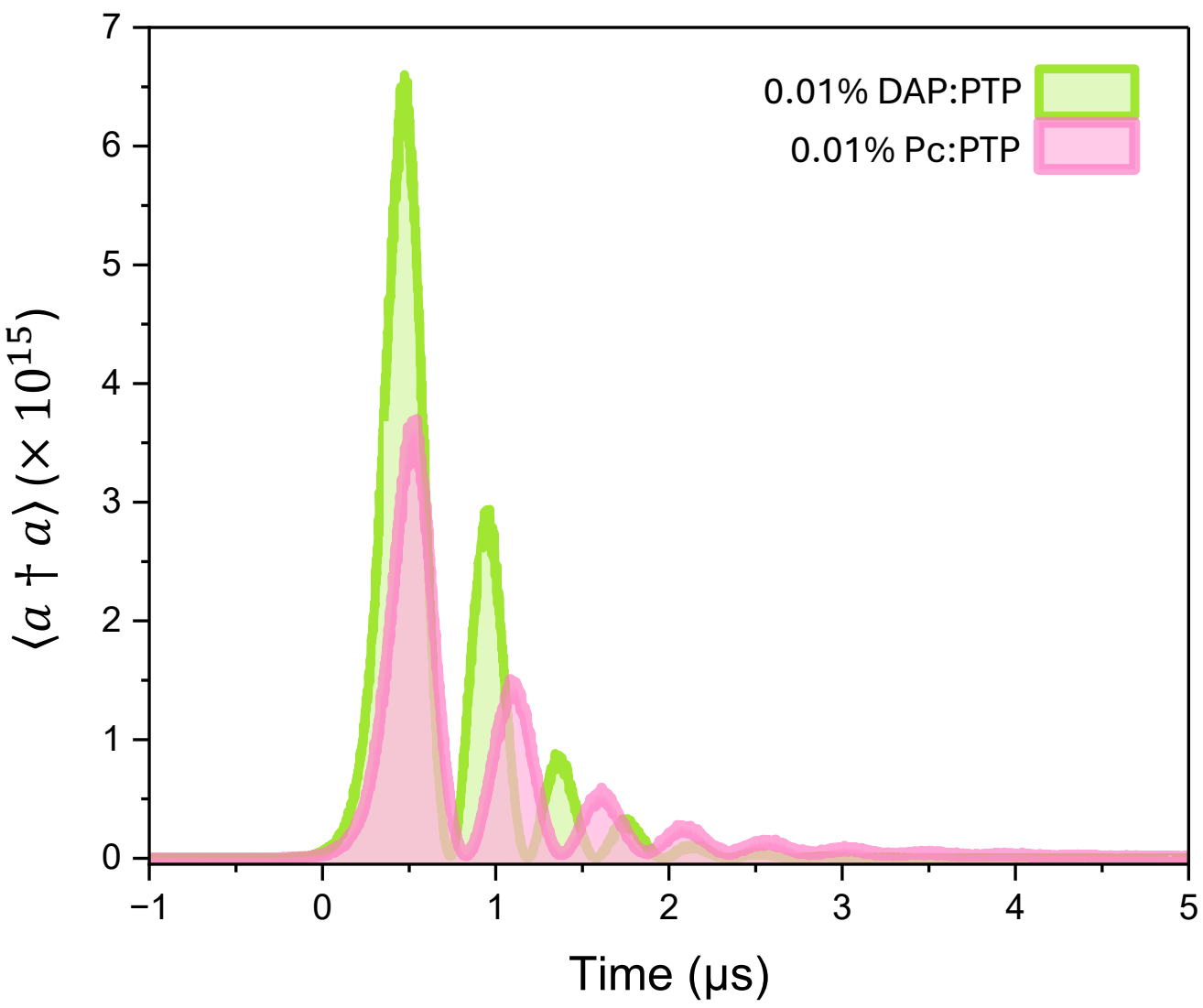

Figure 6: The cavity–spin dynamics were modelled using a set of coupled master equations.

In this framework, the expectation value of the intracavity photon number, $\langle a \dagger a \rangle$ , evolves consistently with the measured time-domain output in mW. The average intracavity photon number $\langle a \dagger a \rangle$ can be estimated using:

$$\langle \boldsymbol{a} \dagger \boldsymbol{a} \rangle = \frac{\boldsymbol{P(t)(1+K)}}{\boldsymbol{h f_{mode} \kappa_c K}} \qquad (3)$$

where $P(t)$ is the peak maser output power, $f_{mode}$ is the cavity resonance frequency, and $K$ is the coupling coefficient (taken as 0.20). The bursts reach peak values of $\langle a \dagger a \rangle \approx 3.7 \times 10^{15}$ photons for Pc:PTP and $\approx 6.6 \times 10^{15}$ photons for DAP:PTP. The photon populations determined from the model are consistent with those shown from the measured microwave output power in Figure 2b. For comparison, Breeze *et al*. (0.053% Pc:PTP, 1.45 GHz), extract and plot $\langle a \dagger a \rangle \approx 0.9 \times 10^{15}$ (their Fig. 3a). Thus, our Pc:PTP and DAP:PTP devices reach 4.1 × and 7.3 × higher intracavity populations, respectively. As $K$ and $f_{mode}$ mode is identical, and $\kappa_c$ is smaller (ours: 0.24 MHz for Pc:PTP and 0.18 MHz for DAP:PTP; $npj =$ 0.18 MHz), the enhancement arises primarily from the increased emitted power.

### 3.5 Thermal Characterisation

Finally, we assessed thermal drift and stability, which set practical limits on linewidth and coherence. Thermal drift of the room-temperature maser system was experimentally characterised in both the gain crystals (Pc:PTP and DAP:PTP) and the STO resonator under optical pumping. Maintaining thermal stability is necessary to avoid cavity resonance frequency drift, which can degrade coherence and overall maser performance. The maser excitation in this work delivers pulsed laser input into the cavity through a machined optical access hole in the cavity wall, simultaneously illuminating the STO resonator and the enclosed gain medium.

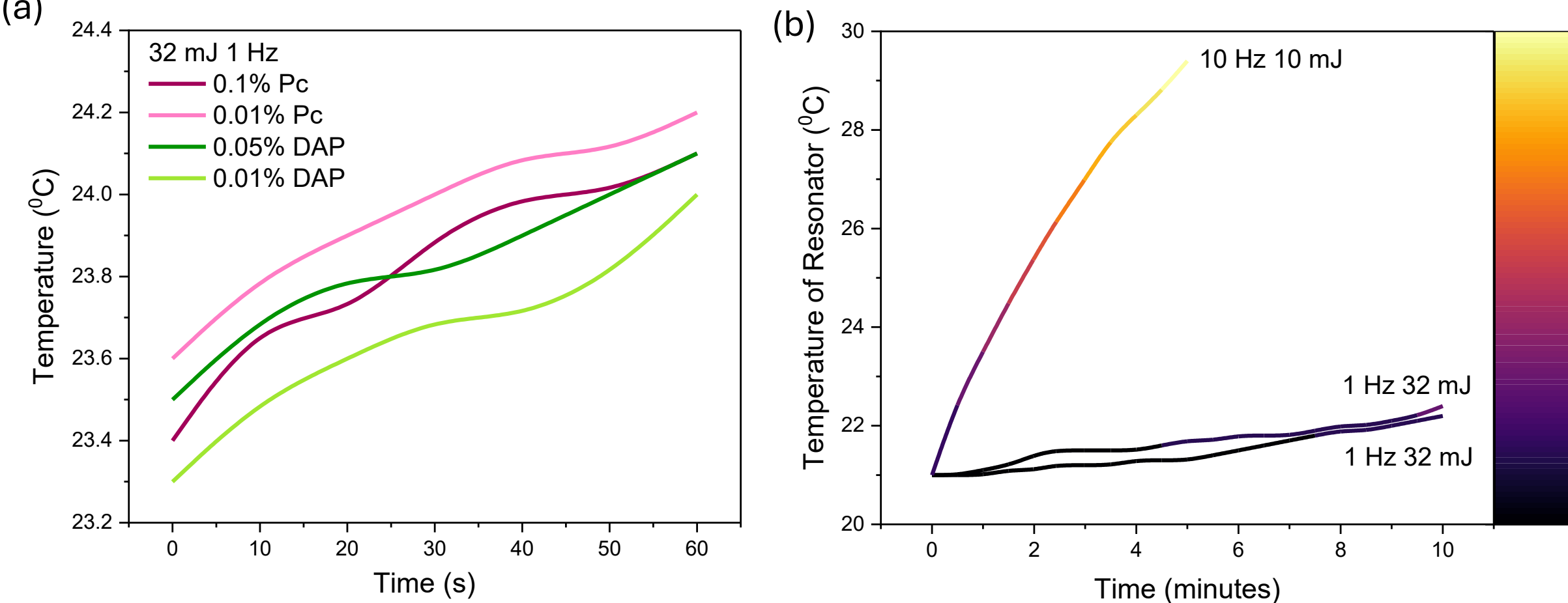

Figure 7: (a) Thermal response of gain media (Pc:PTP and DAP:PTP) and (b) STO resonator under varying laser inputs, reflecting the experimental conditions from the maser output mapping.

Figure 7a illustrates the thermal response of all gain media samples under a 32 mJ, 1 Hz laser input, which corresponds to the maximum optical energy used during the masing output mapping experiments. Despite initial temperature variations, all samples exhibited thermal rise limited to <1°C 60-seconds. Equivalent thermal measurements for the 10 mJ, 1 Hz and 10 mJ, 10 Hz conditions are provided in Figure S4 for reference. Figure 6a shows that the 0.1% Pc:PTP sample exhibited the highest thermal response, with an increase of approximately 1.0 °C after 60 seconds of pumping. In comparison, the remaining samples each show smaller absolute temperature rises of around 0.6 °C under identical conditions. This trend is further supported by higher repetition rate data (10 Hz, 10 mJ; Table 1 row 18), where both 0.01% samples show the smallest absolute increases in temperature (+1.2 °C) compared to +1.5 °C (0.1% Pc:PTP) and +1.9 °C (0.05% DAP:PTP) for higher concentrations.

The relationship between $f_{mode}$ and temperature is given by:

$$f_{mode} = f_0(1 + \alpha \Delta T) \qquad (1)$$

where $f_0$ is the initial frequency, $\alpha$ is the temperature coefficient of the STO resonator, and $\Delta T$ is the temperature change. STO exhibits a temperature coefficient of 170 ppm $K^{-1}$ [27] [28], corresponding to a cavity resonance shift of $\approx$ 0.25 $MHz/°C$, which is significant given the narrow masing linewidth. The enhanced thermal stability of the lower concentration samples significantly reduce temperature-induced frequency drift, directly improving the stability and consistency of masing output. Consequently, these samples demonstrate superior performance compared to higher concentration samples, particularly in continuous or high-duty-cycle operation, where even minor thermal fluctuations can notably degrade maser efficiency and coherence.

Figure 7b presents the thermal response of the STO resonator. The resonator's temperature remain relatively stable under 1 Hz repetition rate with little difference between 10 mJ and 32 mJ laser energies. However, the resonator temperature rises sharply under higher frequency conditions (10 Hz, 10 mJ), increasing by approximately 9°C to reach a final temperature of 30°C within just 5 minutes. This elevated heating arises from both the STO's high temperature coefficient and its exposure to the incident laser path before the gain medium. In addition, at higher repetition rates reduces time for thermal diffusion between pulses leads to cumulative heating, further accelerating the temperature rise.

Table 1 provides a consolidated characterisation of the masing performance across all tested Pc:PTP and DAP:PTP samples. This summary serves as a reference point for evaluating each sample's suitability for different applications, as discussed in the conclusion.

Relative efficiency values are expressed as the slope of maser output power versus pump energy, normalised to the standard 0.1% Pc:PTP sample, which is set as the reference (1.00).

Table 1: Table of characterisation. Comprehensive performance benchmarking of Pc:PTP and DAP:PTP maser media across key quantum and engineering metrics.

| Sample | Pc:PTP 0.1% | Pc:PTP 0.01% | DAP:PTP 0.05% | DAP:PTP 0.01% |
|---|---|---|---|---|
| Loaded Quality Factor of Cavity ($S_{11}$) | ~6000 | | ~8200 | |
| $f_0$ (GHz) | 1.44960 | 1.44949 | 1.47829 | 1.47853 |
| Tuning range (Δf) MHz | 2.05 | 3.15 | 1.67 | 4.28 |
| Max Output (mW) | 0.26 | 1.79 | 0.97 | 2.34 |
| Max Output (dBm) | -5.82 | +2.52 | -0.135 | 3.69 |
| Gradient between 11-32 mJ (μJ/W) | $7.5\times 10^3$ | $4.59\times 10^4$ | $4.03\times 10^4$ | $8.72\times 10^4$ |
| Relative Efficiency (Normalised Gradient Compared to 0.1% Pc:PTP) | 1.00 (Reference) | 6.13 | 5.39 | 11.7 |
| Linewidth at $f_0$ (MHz) (32 mJ laser energy) | 2.98 | 4.99 | 4.30 | 4.81 |
| SNR at $f_0$ (dBc) (32 mJ energy) acquisition bandwidth of 1.44 -1.46 GHz | 22.6 | 29.0 | 28.0 | 28.5 |
| Maser Threshold (mJ) | 4.6 | 2.5 | 6.6 | 5.3 |
| Number of Rabi Cycles at $f_0$ | 5 | 8+ | 6 | 8 |
| Average Rabi Period (ns) | 1010 | 501 | 455 | 400 |

| Normal-mode splitting $\Delta f$ (MHz) | 1.00 | 2.00 | 2.20 | 2.50 |
| --- | --- | --- | --- | --- |
| Error % at $f_0$ | 3.42 | 1.64 | 4.41 | 2.04 |
| Throughput at $f_0$ (Mbps) | 132 | 248 | 208 | 301 |
| Longest Coherence Time $\tau$ (ns) | 390 | 465 | 382 | 297 |
| Sample heating after 30 Seconds 10 Hz 10mJ | +1.5$^{0}$C | +1.2$^{0}$C | +1.9$^{0}$C | +1.2$^{0}$C |

# Discussion

The experimental results from Section 3.2 (Maser Output Mapping) and Section 3.3 (Maser Cavity Frequency Sweep) demonstrate the versatility of the tested maser samples across a range of output powers and linewidth conditions. The observed trends indicate that the choice of operating power and frequency tuning significantly impact application suitability. The maser's output frequency can be tuned by adjusting the cavity plunger, while the output power can be controlled by varying the input optical pump power. This adaptability allows the maser to meet a range of operational requirements. The ability to selectively operate in high-power or low-bandwidth conditions extends the range of potential applications, ensuring that masers remain viable across multiple technology sectors.

Under high optical pump powers, both low-concentration devices surpassed the previous room-temperature maser peak output record of 0.33 mW, with the 0.01% DAP:PTP sample reaching 2.34 mW (+3.69 dBm) and the 0.01% Pc:PTP sample achieving 1.79 mW (+2.52 dBm). This establishes that enhanced maser performance is reproducible across distinct organic gain media, reinforcing the broader applicability of the low-concentration mediums. The efficiency of this power conversion was quantified using a rate analysis $dP/dE = \Delta P/\Delta E$, where the gradient (mW/mJ) provides a measure of how effectively optical pump energy is converted into maser output.

The 0.01% DAP:PTP device exhibit the highest optical-to-microwave conversion efficiency ($8.72 \times 10^4$ µJ/W), more than an order of magnitude greater than the 0.1% Pc:PTP reference ($7.5\times 10^3$ µJ/W), reinforcing the need for targeted gain medium selection. As established in the optical analysis, the markedly higher optical-to-microwave conversion efficiencies observed for the low-concentration samples can be understood in terms of improved pump penetration and reduced reabsorption at lower dopant density. In lower dopants, the pump excites a larger fraction of the crystal volume, enabling more uniform energy deposition, reducing reabsorption and non-radiative losses. These factors account for the 6.1× improvement in 0.01% Pc:PTP and the 11.7× improvement in 0.01% DAP:PTP relative to the 0.1% Pc:PTP reference. The 0.01% samples also exhibited reduced thermal loading under optical excitation, mitigating frequency drift and improving efficiency. These results highlight that low-concentration devices maximise conversion efficiency and possess the stability required to underpin future CW operation.

The lowest maser threshold in this study was measured at 2.5 mJ for 0.01% Pc:PTP, representing a reduction relative to the ~3.2 mJ threshold previously reported for Pc:PTP [19].

For comparison, prior work on DAP:PTP established a threshold of 2.3 mJ when pumped at 620 nm [20]. In our study, the 0.01% DAP:PTP sample yielded a higher threshold of 5.3 mJ under 532 nm excitation, which is attributable to the non-optimal pump wavelength rather than an intrinsic material limitation, given that the same crystal exhibited lower thresholds under 620 nm pumping in earlier work.

Low-concentration Pc:PTP and DAP:PTP samples yielded SNR values of 28-29 dBc, highlighting a pronounced noise-reduction benefit. Our 0.1% Pc:PTP device (22.6 dBc) closely matches the ~20 dBc previously reported for the same concentration, confirming that reduced dopant concentration directly enhances output stability. Such improvements are accumulative; they elevate organic masers into a noise-resilient performance domain directly relevant to high-impact applications. In quantum sensing, higher SNR translates to greater sensitivity and precision. In secure communication, it enhances channel fidelity. In deep-space tracking and VLBI astronomy, detection limits are improved under noise-dominated conditions. The observation that three independent samples achieve this level of stability underscores that the stability is reproducible and platform-level, not specific to a single gain medium.

Throughput analysis revealed peak values up to 301 Mbps for 0.01% DAP:PTP, compared with 248 Mbps for 0.01% Pc:PTP and lower values in higher-concentration samples. Together with the observed number of Rabi cycles (up to 8 in both 0.01% Pc:PTP and 0.01% DAP:PTP), the results demonstrate that low-concentration devices sustain strong coherent dynamics while maintaining high signal quality. Further analysis showed the longest coherence time of 465 ns in the 0.01% Pc:PTP sample, contrasting with shorter values in higher concentrations, including the 0.1% Pc:PTP device, which consistently performed the weakest. While the 0.05% DAP:PTP sample exhibited longer coherence than the 0.1% Pc:PTP case, its response was less repeatable and displayed anomalous linewidth behaviour, as detailed in the Supplementary Information.

Lower-power masers exhibit extended temporal coherence times and narrower linewidths, making them more relevant for applications requiring long-term signal stability and phase coherence, highlighting the inherent trade-off between high-power masing for strong signal amplification and long coherence times for precise spectral purity. The narrow-bandwidth regime, characterised by threshold-limited behaviour and frequency tuning away from $f_0$ (Section 3.3), linewidths approached the kilohertz regime. Under these conditions, the maser demonstrates longer coherence times (~300-465 ns) and coherence lengths nearing 150 m, suggesting future applicability in low-phase-noise applications, deep-space tracking, and Very-long-baseline interferometry (VLBI) radio astronomy, where frequency stability is more critical than output power.

Thermal stability is a critical determinant of maser performance under extended or high-duty-cycle operation, where cumulative heating drives frequency drift and destabilises output. Figure 6 highlights two key aspects: first, the resonator exhibits sharp thermal loading at elevated repetition rates (10 Hz), underscoring the importance of managing optical heating for future CW studies. Second, the gain media show a strong concentration dependence. Both low-concentration samples (0.01% Pc:PTP and 0.01% DAP:PTP) exhibited the lowest absolute heating (+1.2 °C after 30 s at 10 Hz, 10 mJ), compared with +1.5 °C for 0.1% Pc:PTP and +1.9 °C for 0.05% DAP:PTP. The anomalously high heating in the 0.05% DAP:PTP likely contributes to its irregular linewidth response. This stability contrasts with the robust stability observed in the 0.01% DAP:PTP device. These findings establish that lower dopant concentrations mitigate thermal loading and frequency drift, providing a clear optimisation route towards CW maser operation.

## Conclusion and Outlook

This study addresses a key gap in maser research by providing the first systematic characterisation of multiple L-band masing samples under consistent experimental conditions. While prior work has often focused on individual gain media or isolated performance criteria, a comparative understanding of material behaviour under matched conditions has yet to be addressed. This limits the ability to optimise masers for specific applications. By establishing baseline performance metrics, including output power, gradient efficiency, tuning range, SNR, and coherence properties, this work lays the foundation for targeted material and concentration selection.
These performance benchmarks underscore the feasibility of advancing toward continuous-wave L-band masers, where thermal management and signal stability will be conclusive. They also highlight the immediate relevance of low-concentration organic systems for practical applications, from noise-resilient sensing to secure communications.

Further advancements in resonator design and active signal stabilisation to achieve narrower linewidths will expand maser applicability to ultra-precise timing standards, secure communications, quantum logic operations, and deep-space Doppler tracking.

By systematically characterising material performance across key parameters, this work provides guidance for optimising masers for high-impact applications. These findings represent a pivotal step in bridging the gap between fundamental maser physics and real-world deployment and highlight pathways to tailor device performance for integration in quantum information processing and secure L-band communication networks.

## Methods

*Theoretical throughput*

The theoretical throughput of the maser signal was estimated using the Shannon capacity formula, which describes the maximum achievable data rate for a communication channel in the presence of noise [29]. This was applied to the 0.01% DAP:PTP sample, using:

$$Capacity = B\log_2(1 + \frac{C}{N}) \quad (2)$$

Where $B$ = Bandwidth (9.702 MHz) is the effective resolution bandwidth of the FFT spectrum, $C$ = Signal Power (2.23 mW) is the measured maser signal power, $N$ = Noise Power (1 pW, from from -90 dBu) is the noise power derived from the calibrated noise floor of –90 dBu in a 50 Ω load. The Shannon formula assumes an additive white Gaussian noise (AWGN) channel.

$$Capacity = 9.702 \times 10^6 \log_2(1 + \frac{2.23 \times 10^{-3}}{1.00 \times 10^{-12}})$$

$$Capacity = 301.29\ Mbps$$

This high theoretical capacity reflects the narrow linewidth and high signal intensity of the masing output. While not optimised here for digital modulation, it is a valuable benchmark for evaluating the information-carrying potential of coherent microwave sources compared to conventional classical systems. Although representing an idealised upper bound, it underscores the promise of maser-based RF systems for high-fidelity, quantum-compatible data transmission frameworks.

*SNR*

SNR values were obtained directly from the PicoScope 6428E FFT analysis. According to the PicoScope User Manual [30] SNR is defined as:

$$\boldsymbol{SNR = 20log_{10}\left(\frac{RMS\ value\ of\ carrier\ ('datum')}{\sqrt{\sum of\ squares\ of\ all\ values\ excluding\ datum\ and\ harmonics}}\right)} \quad (3)$$

where the ‘datum’ corresponds in this work to the carrier peak at the masing transition frequency.

The spectral coherence time for each measurement is calculated via:

$$\tau_c = \frac{1}{\pi\Delta f} \quad (4)$$

Where Tc is coherence time (ns), and $\Delta f$ is linewidth (MHz). The spectral coherence length, the physical distance of where a wave maintains its phase coherence, has the formula:

$$L_c = c \times \tau_c = \frac{c}{\pi\Delta f} \quad (5)$$

Where $L_c$ is the spectral coherence length (m), and c is the speed of light in free space.

*Intracavity Photon Number*

The RF envelope was obtained with the Hilbert transform $V_{pk}(t) = |hilbert(v(t))|$. The cycle-averaged (RMS) power is:

$$\boldsymbol{P(t)} = \frac{\boldsymbol{V_{pk}(t)^2}}{\boldsymbol{2 \cdot 50\,\Omega}} \qquad (6)$$

The intracavity photon number then follows:

$$\langle \boldsymbol{a} \dagger \boldsymbol{a} \rangle = \frac{\boldsymbol{P(t)(1+K)}}{\boldsymbol{hf_{mode}\kappa_c K}} \qquad (7)$$

with $K = 0.20$, $f_{mode} = \{1.449, 1.478\}$ GHz, and $\kappa_c/2\pi = \{0.24{,}0.18\}$ MHz. With these parameters, the burst peak at $3.7 \times 10^{15}$ (0.01% Pc:PTP) and $6.6 \times 10^{15}$ (0.01% DAP:PTP) photons. For a cross check, the instantaneous power $p(t) = v(t)^2/R$, with Pc:PTP = 1.79 mW, and DAP:PTP = 2.34 mW, the peak photon number becomes $4.4 \times 10^{15}$ (Pc:PTP) and $1.3 \times 10^{16}$ (DAP:PTP).

*Estimation of penetration depth*

The Beer–Lambert formalism is well established in absorption spectroscopy, relating absorbance to material properties via $A = \varepsilon(\lambda)cd$, where A is the absorbance at wavelength $\lambda$, $\varepsilon(\lambda)$ is the molar extinction coefficient (L mol$^{-1}$ cm$^{-1}$), $c$ is the concentration in (mol L$^{-1}$), and $d$ is the optical path length (cm). This can also be expressed as $I = I_0 e^{-\alpha d}$, where $I$ and $I_0$ are the transmitted and incident intensities, respectively, and $\alpha$ is the absorption coefficient in (cm$^{-1}$).

In terms of optical depth, the Beer–Lambert law relates the penetration depth $\delta$ to the transmitted and absorbed fractions of light. In the present work, we use reported penetration depths $\delta = 1/\alpha$ reported in [24].

For a pump path length $L_{pump}$, this gives the dimensionless form:

$$\tau = \frac{L_{pump}}{\delta} \qquad (8)$$

where $L_{pump} = 3\,mm$ is the optical pump path length (corresponding to the transverse dimension of the gain medium).

So, for 0.1 mol % dopant, Takeda reports $\delta = 0.7 - 1.0\,mm$. Substituting into the eq. (3) gives:

$$\boldsymbol{\tau} = \frac{\boldsymbol{3\,mm}}{\boldsymbol{0.7 - 1.0\,mm}} = \boldsymbol{4.3 - 3.0}$$

corresponding to:

$$A = 1 - e^{-\tau} \approx 95 - 98.6 \text{ absorbed,}$$
$$T = e^{-\tau} \approx 1.4 - 5.0\,\% \text{ transmitted}$$

Thus, a high concentration of Pc:PTP absorbs nearly all incident pump light within the first millimetre of the crystal.

For 0.01 mol % dopant, Takeda reports $\delta \gtrsim 2\,mm$. Mapping this to the same 3 mm path gives

$$\tau = \frac{3\ mm}{2.0\ mm} = 1.5$$

which corresponds to $A \leq 77.7\%$ absorbed, $T \geq 22.3\ \%$ transmitted. This reduced absorption enables significantly deeper penetration of the 532 nm pump light, leading to more uniform excitation across the crystal bulk.

*Time-domain Rabi frequencies*

Time-domain Rabi frequencys $f_R$ were obtained from the inverse of the envelope peak-to-peak spacing in the transient maser oscillations, $f_R = 1/T_{peak}[ns]$ reported in MHz.

*Purcell Factor*

The efficiency of spin–photon coupling in maser systems is strongly influenced by the cavity's magnetic Purcell factor $F_m = \frac{4\pi}{3} \cdot \frac{Q}{V_m}$, which captures the enhancement of magnetic dipole transitions due to resonant field confinement [13], [26]. In this work, we employ cylindrical annular cavities fabricated from strontium titanate ($SrTiO_3$, also STO), a high-permittivity dielectric with $\varepsilon_r \approx 318$ at room temperature, supporting $TE_{01\delta}$ characterised by a strong axial magnetic field component aligned with the crystal axis, ensuring maximal overlap with the spin ensemble. For cavities containing Pc:PTP and DAP:PTP, the measured loaded quality factors at the resonant frequency were approximately 6000 and 8200, respectively. These high-Q factors, combined with the small effective magnetic mode volume provided by the STO geometry $V_m \approx 0.25\ cm^3$, result in magnetic Purcell factors on the order of $10^8$, indicating significantly enhanced spin–photon coupling [13].

This enhancement is central to achieving the strong coupling regime, where coherent energy exchange between the spin ensemble and cavity photons occurs faster than their respective decay rates [31].

*Experimental setup*

The maser cavity was designed using a commercially available finite-difference time-domain (FDTD) simulator (CST Studio Suite®). The cavity was designed to support the $TE_{01\delta}$ mode, which is coupled to the maser signal via the coaxial loop antenna. The loaded maser cavity consists of a cylindrical copper cavity with an inner diameter of 40 mm and a height of 35 mm, housing a STO cylindrical annulus resonator. The STO resonator, is positioned centrally within the cavity on a 2 mm polytetrafluoroethylene (PTFE) stand. The resonator has an outer diameter of 10.12 mm, an inner diameter of 3.05 mm, and a height of 14.52 mm.

Positioned at the centre of the STO annulus is a single organic crystal ~3 mm in diameter and approximately 5-7 mm in length, serving as the gain medium (either Pc:PTP or DAP:PTP). This resonator-crystal assembly is mounted on a PTFE stand, elevating it by 2 mm from the cavity base to optimise field uniformity and coupling.

Stimulated microwave emission is extracted via a loop antenna connected to the end of a coaxial feed, serving as the cavity's output port. The fundamental resonance frequency $f_0$ is tuned manually using a mechanical plunger and was aligned to the masing transition for each medium: 1.449 GHz for Pc:PTP and 1.478 GHz for DAP:PTP.

No external amplification, filtering, or signal processing components were introduced between the maser cavity and the digital oscilloscope, ensuring that all recorded waveforms reflect the intrinsic emission characteristics of the masing process. This direct detection approach also helps validate the assumption that the transmission circuit remains impedance-matched at 50 Ω, which is standard for RF and microwave systems. Accordingly, output power $P$ was calculated from the measured voltage signal using the relation $P = {V^2}/{R}$, where $V$ is the peak maser signal voltage ($V_{peak}$) obtained directly from the oscilloscope trace, and $\Omega = 50$ is the assumed load impedance. For example, a peak voltage of 342.44 mV corresponds to a power output of 2.34 mW (or +3.69 dBm). As the oscilloscope input was terminated at 50 Ω and no external gain or loss elements were present, these voltage-to-power conversions represent absolute, calibrated output powers.

*Optical pump*

Optical excitation of the organic masing samples was achieved using a 532 nm Nano S 120-20 pulsed laser, delivering ~5 ns nominal pulses across a tuneable energy range (2.5 mJ to 32 mJ, measured at laser output). The laser beam, with a spot diameter of 2.5 mm, was directed perpendicularly onto the surface of the organic single crystals positioned within the STO dielectric resonator.

This excitation initiates population inversion within the triplet manifold, enabling coherent microwave amplification within the cavity. The stimulated emission process within the maser cavity leads to microwave amplification at 1.449 GHz (Pc:PTP) and 1.478 GHz (DAP:PTP), between the $|X\rangle$ and $|Z\rangle$ triplet sublevels. Key parameters are the intersystem crossing rate ($k_{ISC}$), and the internal conversion rate ($k_{IC}$). Spin-lattice relaxation rates ($\Upsilon_{xz}, \Upsilon_{xy}$, and $\Upsilon_{yz}$,) govern transitions between $T_1$ sublevels, with decay rates ($k_t$) facilitating the return of $T_1$ sublevels to the singlet ground state, $S_0$. Collectively, these transitions complete the masing cycle

This redshift behaviour can be explained by considering the electronic structure of DAP:PTP. Introducing two nitrogen atoms along the aromatic backbone impacts the molecular orbital energies, leading to a redshift in the absorption spectrum. As illustrated in the Jablonski diagram (Figure 1b), this DAP:PTP redshift is primarily attributed to the influence of nitrogen's higher electronegativity relative to carbon on the Pauling scale (3.04 vs. 2.55, respectively). The nitrogen atoms, when introduced into the aromatic backbone of DAP:PTP, draw electron density towards themselves, exerting an electron-withdrawing effect. This interaction selectively lowers the energy of the Lowest Unoccupied Molecular Orbital (LUMO), which is more spatially extended and sensitive to substituent effects, while the Highest Occupied Molecular Orbital (HOMO), being more localised and stabilised by the aromatic π-system, remains largely unaffected. As a result, the HOMO–LUMO gap narrows, shifting absorption towards longer wavelengths.

*Thermal Camera*

An infrared camera (SEFRAM 9832, Sefram Instruments, St Etienne, France) was used in the experiments. The camera was calibrated with an emissivity setting of 0.93, matching the estimated emissivity of strontium titanate (STO), the material under test, which typically ranges between approximately 0.90 and 0.96 [32] due to its glass-like surface properties; as a result, the potential temperature correction is negligible (≤ 0.2°C), ensuring accurate thermal measurements without the need for further adjustment.

*The Photon build-up equation*

The enhanced coupling observed in DAP:PTP can also be understood by examining the photon gain dynamics through the rate-equation framework. The photon build-up equation,

$$\frac{d\bar{n}}{dt} = -\omega_{mode}\left[\frac{1}{Q_0}(\bar{n} - \alpha_T T_0) + \frac{1}{Q_e}(\bar{n} - \alpha_T T_e)\right] + B_{XZ}(N_x - N_z)(\bar{n} - \alpha_T T_s) \quad (9)$$

where $\bar{n}$ is the average photon number in the maser mode, $\omega_{mode}$ is angular frequency of the maser cavity mode, $Q_0$ is intrinsic (unloaded) quality factor of the cavity, $Q_e$ is coupling quality factor to the external circuit or load, $\alpha_T$ is thermal photon occupation coefficient, $T_0$ is ambient temperature, $T_e$ is effective external load temperature, $T_s$ effective spin temperature, $B_{XZ}$ is Einstein B coefficient for stimulated emission between triplet sublevels $|X\rangle$ and $|Z\rangle$, $N_x - N_z$ represents the net population difference between these sublevels, proportional to the number density of spin-polarised molecules actively contributing to masing.

Equation 9 describes how the intracavity photon population $\bar{n}$ evolves in time, incorporating both cavity loss and spin-mediated amplification via the term $B_{XZ}(N_x - N_z)$. This gain term includes contributions from both the effective spin-photon coupling and the population difference between the triplet sublevels involved in masing. While the cooperativity $C^*$ captures the strength of spin–photon coupling, it does not fully describe the efficiency of population inversion. The significantly higher photon populations and masing output observed in DAP:PTP suggest that this material supports a more efficient redistribution among triplet sublevels, enabling stronger masing gain via a larger effective value of $N_x - N_z$.

*Crystal Growth*

Crystals of Pc:PTP and DAP:PTP were grown using the Bridgmann technique described previously [20]. Doped powders of p-terphenyl (extensively zone refined prior to use) were made by grinding in a pestle and mortar the corresponding molar ratios of pentacene (0.1%, 0.01%) and DAP:PTP (0.01%, 0.05%) with p-terphenyl. The powders were each decanted into 3 mm (ID) borosilicate tubes (Hilgenberg GmbH) and sealed under argon. The powders were then passed through a furnace heated to 216°C at 4 mm/hr to yield large cylindrical single crystals.